\documentclass[a4paper,fleqn,usenatbib]{mnras}

\usepackage{newtxtext,newtxmath}

\usepackage[T1]{fontenc}
\usepackage{ae,aecompl}

\usepackage{graphicx}	
\usepackage{amsmath}	
\usepackage{amssymb}	

\title[Turbulent Vertical Mixing]{Turbulent Vertical Mixing in Hot Exoplanet Atmospheres}

\author[K. Menou]{
Kristen Menou,$^{1, 2}$
\\
$^{1}$  Physics \& Astrophysics Group, Dept.   of  Physical  \&  Environmental  Sciences,  University  of  Toronto  Scarborough,\\   1265 Military Trail, Toronto, Ontario, M1C 1A4, Canada \\
$^{2}$ Dept.  of Astronomy \& Astrophysics, University of Toronto.
50 St.  George Street, Toronto, Ontario, M5S 3H4, Canada \\
}

\date{Accepted XXX. Received YYY; in original form ZZZ}

\pubyear{2015}

\begin{document}
\label{firstpage}
\pagerange{\pageref{firstpage}--\pageref{lastpage}}
\maketitle

\begin{abstract}
Turbulent vertical transport driven by double-diffusive shear instabilities is identified as likely important in hot exoplanet atmospheres. In hot Jupiter atmospheres, the resulting vertical mixing appears sufficient to alleviate the nightside cold trap, thus facilitating the maintenance of nocturnal clouds on these planets. The strong level of vertical mixing expected near hot Jupiter thermal photospheres will impact their atmospheric chemistry and even their vertical structures where cloud radiative feedback proves important.
\end{abstract}

\begin{keywords}
hydrodynamics -- radiative transfer  -- planets and satellites: atmospheres -- turbulence -- astrochemistry -- diffusion 
\end{keywords}



\section{Introduction}
Observational campaigns with the radial velocity and transit detection methods have led to the discovery of a vast and diverse population of hot exoplanets. Chief among them, hot Jupiters are the best characterized type of exoplanets. Hot Jupiters have become de-facto testing grounds for many efforts to interpret atmospheric data from exoplanets, using measurements of secondary eclipses, phase curves and transmission spectra \citep{2007arXiv0706.1047C, 2010ARA&A..48..631S,2010RPPh...73a6901B, 2016SSRv..205..285M}.

Hot Jupiter atmospheres have revealed or hinted at the existence of many of the phenomena known to occur in the atmospheres of solar system planets, such as radiatively driven circulation, atmospheric chemistry, condensation and clouds. However, hot Jupiters are also expected to exhibit new phenomena in relation to their unique physical properties as hot and tidally-locked gas giants. For example, supersonic winds may prevail in these strongly-forced atmospheres even in the presence of dissipation through shocks \citep{2010ApJ...725.1146L, 2013MNRAS.435.3159D, 2016A&A...591A.144F}. Furthermore, magnetic drag and associated ohmic dissipation likely operate and are often invoked to explain the radius inflation exhibited by many hot Jupiters \citep{2010ApJ...714L.238B, 2010ApJ...719.1421P, 2012ApJ...745..138M, 2018AJ....155..214T}.

{One aspect of atmospheric dynamics that is challenging to capture and often treated on a planet-specific basis is the magnitude of vertical mixing present  in a planetary atmosphere that acts to prevent gravitational settling of condensates. The process is often modeled as an eddy diffusivity, $K_{\rm zz}$, with vastly different profiles adopted by modelers of both Solar System and exoplanet atmospheres \citep[see Figure~1 of][]{2018ApJ...866....1Z}. In the exoplanet context, \cite{2013A&A...558A..91P} have advocated a $K_{\rm zz}$ profile, computed from the global circulation pattern in a simulation of the hot Jupiter HD209458b, which significantly revises the previously adopted profile by \cite{2011ApJ...737...15M} for the same planet. Recently, \cite{2018ApJ...866....1Z, 2018ApJ...866....2Z} have developed a general theory for global mean vertical tracer mixing in planetary atmospheres. Additionally, it is important to consider the possibility that instabilities driven by vertical shear could also play a significant role for vertical mixing. This process must be elucidated outside of general circulation models since, by construction, the primitive equations of meteorology filter out the vertical motions at the origin of shear instabilities.  }

In this letter, we introduce a new physical process that may prove important to address the hot Jupiter phenomenology: enhanced turbulent vertical transport from double-diffusive shear instabilities. This class of instabilities, which has its origins in the stellar astrophysics literature,  requires a hot and opaque gaseous medium to operate. To the best of our knowledge, it is not thought to operate in colder Solar System atmospheres, which makes this process specifically relevant to hot exoplanets. While the instability has been discussed briefly before by \cite{2010ApJ...725.1146L} in the context of local idealized fluid simulations, we adopt here a more global view and suggest that it could be key in establishing strong vertical mixing in hot Jupiter atmospheres and, by extension, other hot exoplanet atmospheres.

\section{Double-Diffusive Shear Instability}

Instabilities in an atmospheric flow can be driven by vertical shear $S \equiv \partial V_H / \partial z$ where  $V_H$ is the horizontal velocity in the rotating frame of the planet and $z$ is the vertical coordinate.  Often, this type of instability is inhibited by a strongly stabilizing stratification, whose magnitude is measured by the Brunt-Vaissala frequency $N$:
\begin{equation}
N^2 = \frac{g}{\gamma} \frac{d}{dz} \ln \frac{P}{\rho ^ \gamma} \rightarrow \frac{\gamma -1}{\gamma} \frac{g}{H_p}
\end{equation}
where $H_p=\cal{R}$ $T/g$ is the pressure scale height, the gas is assumed to behave ideally, $\cal{R}$ is the gas constant, $\gamma$ is the adiabatic index and $g$ is the gravitational acceleration. It is well known \citep{1961hhs..book.....C} that adiabatic perturbations are stable if the Richardson number of the shear flow
\begin{equation}
Ri \equiv \frac{N^2}{S^2} \geq \frac{1}{4}.
\end{equation}
As we shall see below, hot Jupiter atmospheres are typically expected to satisfy this adiabatic stability criterion and thus be Richardson-stable.

In the presence of thermal diffusivity, however, the stable vertical stratification can be effectively weakened to the point that small scale unstable modes can grow. The destabilizing role of thermal diffusivity is well known in the context of stellar astrophysics \citep[e.g. leading to the so-called GSF instability for unstable rotation profiles][]{1967ApJ...150..571G,1968ZA.....68..317F,2004ApJ...607..564M}.\footnote{We find that hot Jupiter atmospheric flows are likely rotationally stable \citep[see also][]{2010ApJ...725.1146L}} 
Radiative diffusion from an ample bath of photons in an opaque hot gas is the key factor facilitating this class of (double-)diffusive instabilities in the stellar context. The same logic applies to hot exoplanet atmospheres, as we shall illustrate with hot Jupiters below. By contrast, cool atmospheres such as those of Solar System planets will not be affected by the diffusive shear instabilities of interest here.
 
In the presence of a strong enough (radiative) thermal diffusivity
\begin{equation}
\xi \equiv \frac{16}{3} \frac{\gamma -1}{\gamma} \frac{\sigma T^4}{\kappa \rho p},
\end{equation}
obtained in the radiative diffusion limit, and a weak enough kinematic viscosity $\nu$, for the opaque atmospheric gas, a secular version of the Richardson criterion becomes relevant. The criterion for double-diffusive instability is \citep{1992A&A...265..115Z, 2015ApJ...801..137P}
\begin{equation}
Ri_s \equiv  \frac{\nu}{\xi} Re_c \frac{N^2}{S^2} < \frac{1}{4},
\end{equation}
where $Re_c \simeq 1000$ is the critical Reynolds number for turbulence, with instability favoured for $\nu / \xi \ll 1$. Hot exoplanet atmospheres, in particular hot Jupiters, are such that their opaque regions with large $\xi$ and small $\nu$ values can become subject to double-diffusive shear instabilities. In the presence of such instabilities, turbulent vertical transport is expected to occur with a strength captured by the effective viscosity
 \begin{equation}
\nu_{\rm eff} \simeq  \frac{S}{N} \xi
\end{equation}
with a limiting strength set by $\xi / N \ll H_p^2$ (to guarantee unstable wave-modes smaller than one atmospheric scale height). As long as the gas kinematic viscosity $\nu \ll  \xi$, viscosity has no direct effect on the magnitude of the turbulent transport in this formulation.

\section{Scalings for Hot Exoplanet Atmospheres} 
 
Let us now consider the range of unstable regimes and turbulent viscosities that can be expected over the vast parameter space of hot exoplanet atmospheres. For concreteness, we fix the atmospheric surface gravity and opacity coefficient to reference values $g_0$ and $\kappa_0$ respectively. Gas is assumed to behave ideally. We write the gas kinematic viscosity $\nu \simeq \lambda c_s$, where $\lambda$ is the gas mean free path with a fixed collisional cross-section (assumed $\sim 3 \times 10^{-16}$~cm$^2$ as for H gas, for simplicity) and 
\begin{equation}
c_s = \sqrt{{\cal R}  T}
\end{equation}
is the (isothermal) sound speed. 

Under these conditions, we derive the leading-order scalings for $\nu$, $\xi$, $N$ and $S$ in terms of atmospheric pressure, $P$, and temperature, $T$,  for given values of    $g_0$ and $\kappa_0$. We find
\begin{eqnarray}
\nu & \propto & \frac{T^{3/2}}{P}, \\
\xi & \propto & \frac{T^5}{\kappa_0 P}, \\
N & \propto & \frac{g_0}{T^{1/2}},\\
S & \propto & \frac{g_0}{T}.
\end{eqnarray}

As a result, the secular Richardson number and turbulent viscosity obey, to leading-order,

 \begin{eqnarray}
Ri_s & \propto & \frac{\kappa_0 P}{T^{5/2}},\\
\nu_{\rm eff} & \propto & \frac{T^{9/2}}{\kappa_0 P^2},
\end{eqnarray}
with no explicit dependence on $g_0$. These idealized scalings show that an unstable  atmosphere is more vigorously stirred by diffusive shear instabilities at shallower depths (low $P$) and higher temperatures.  

It is useful at this point to evaluate the degree of stability of a prototypical atmosphere, resembling that of the hot Jupiter HD209458b. We adopt the following representative values to ground our discussion: $T = 1500$K, $P =1$bar, $N/S \simeq 5$ and $Ri_s \simeq 10^{-4} N^2/S^2$, with $g = 25$m~s$^{-2}$ and $H_p=10^6$m. Our estimate of the shear rate $S$ is based on a velocity differential of $1800$m~s$^{-1}$ spread over two pressure scale heights, as inferred from the core equatorial jet in a typical simulated zonal wind profile \citep[e.g., Fig.~3 in][]{2012ApJ...750...96R}. Note that these values correspond to $Ri =25$ so that the atmosphere is indeed Richardson-stable despite the presence of strong vertical shear.

Once the role of diffusion is accounted for, however, we find ample room for double-diffusive instability, with $Ri_s \simeq 2 \times 10^{-3} \ll 1/4$.  The turbulent viscosity coefficient evaluates to $\nu_{\rm eff} \simeq 10^9 P_{\rm bar}^{-2}$cm$^2$s$^{-1}$, which is comparable to or greater than other values reported in the literature for $K_{\rm zz}$  above the 1 bar pressure level \citep[e.g.][]{2013A&A...558A..91P}. The turbulent vertical transport becomes unrealistically strong as one approaches the thermal photosphere\footnote{{We refer to the thermal photosphere as the pressure level above which thermal photons freely escape to space. In the grey-radiation models presented in \S~4, this corresponds to the 50 mbar level approximately.}} according to this idealized prescription (with $K_{\rm zz} \sim 10^{13}$cm$^2$s$^{-1}$ at 10 mbar) and it is likely limited before that happens (see simulation results below). The transport is also expected to suddenly cut off above the photosphere as heat transport ceases to operate in the radiative diffusion approximation (our working assumption in writing Eq.~3 for $\xi$) and photons rapidly become free to stream out to space in the optically-thin component of the atmosphere.

Combining the specific results for HD209458b with the scalings across the hot exoplanet parameter space above, we infer that diffusive shear instabilities may be operating over a vast parameter space for hot exoplanets, down to warm effective temperatures perhaps as low as $500$K. The magnitude of the resulting transport is strongly dependent on atmospheric temperatures, however, so that the strongest effects would typically be expected on the hottest planets.

\section{Global Circulation Model for HD209458b}

Having established the relevance of double-diffusive shear instabilities for hot exoplanet atmospheres, we move on to an illustration of their consequences through the lens of a numerical simulation for a specific hot Jupiter, HD209458b. One of the key motivations for this numerical exercise is that shear instabilities are sensitive to details of the atmospheric flow. In addition, a strong enough turbulent vertical transport can feedback on the circulation pattern (i.e., reduce the shear) in such a way that the instability ends up operating at a reduced level or even marginally. 
 
Performing a fully-coupled atmospheric flow simulation with a prescription for turbulent vertical transport that is dynamically-adjusted with the flow is beyond the scope of this study. Instead, we settle for the next best approach, which is the specification of a static imposed vertical transport profile, from which we can run diagnostics on the level of turbulent vertical transport that may be expected in a more realistic fully-dynamic simulation.

We have established above that a reasonable profile of turbulent vertical diffusion for HD209458b is :
\begin{equation}
K_{zz} \simeq 10^9 \left( \frac{P}{1~{\rm bar}} \right)^{-2}~{\rm cm^2 ~s^{-1}}. 
\label{eq:transport}
\end{equation} 
We impose this profile for the mixing of momentum and entropy in an HD209458b simulation using the built-in vertical transport scheme in our meteorological solver, IGCM3 \citep{2012ApJ...750...96R}. It solves turbulent vertical transport of a quantity ($Q$) as a diffusion problem 
 \begin{equation}
\frac{\partial Q}{\partial t} = \frac{\sigma}{H_p^2}  \frac{\partial }{\partial \sigma} \left( \sigma K_{zz} \frac{\partial Q}{\partial \sigma} \right),
\end{equation} 
where $\sigma=P/P_{\rm bot}$ is the pressure coordinate (scaled to the bottom pressure $P_{\rm bot}$) and $K_{zz}$ is the diffusion coefficient for quantity $Q$. Zero flux boundary conditions are imposed. Viscous heating is not returned to the flow in this implementation, while it could prove important where friction is strong.

Our simulation setup is in most ways similar to that in \cite{2012ApJ...750...96R}, with the following exceptions. Our model is truncated at the 10 mbar pressure level, which approximately corresponds to the thermal photosphere for this model. A model resolution of T21L30 (64x16 with 30 vertical levels) is adopted.  Broadly consistent results are obtained at a lower resolution of T10L20 (32x8 with 20 vertical levels). The model is integrated forward in time for 100 planet rotation periods, rather than 2000 as in \cite{2012ApJ...750...96R}, as dictated by computational limitations: we must adopt small time-steps in the presence of the strong vertical transport imposed ($\Delta t$ must be reduced by a factor of 50 in the run with vertical transport, for numerical stability). Nevertheless, this run time is sufficient to establish a steady wind pattern in the upper simulation domain (see Figure 1). The deepest atmospheric levels (below $\sim 1$~bar) require much more time for spin-up but they are also largely irrelevant to our discussion of vertical transport above the 1 bar level.

\begin{figure}
	\includegraphics[width=\columnwidth]{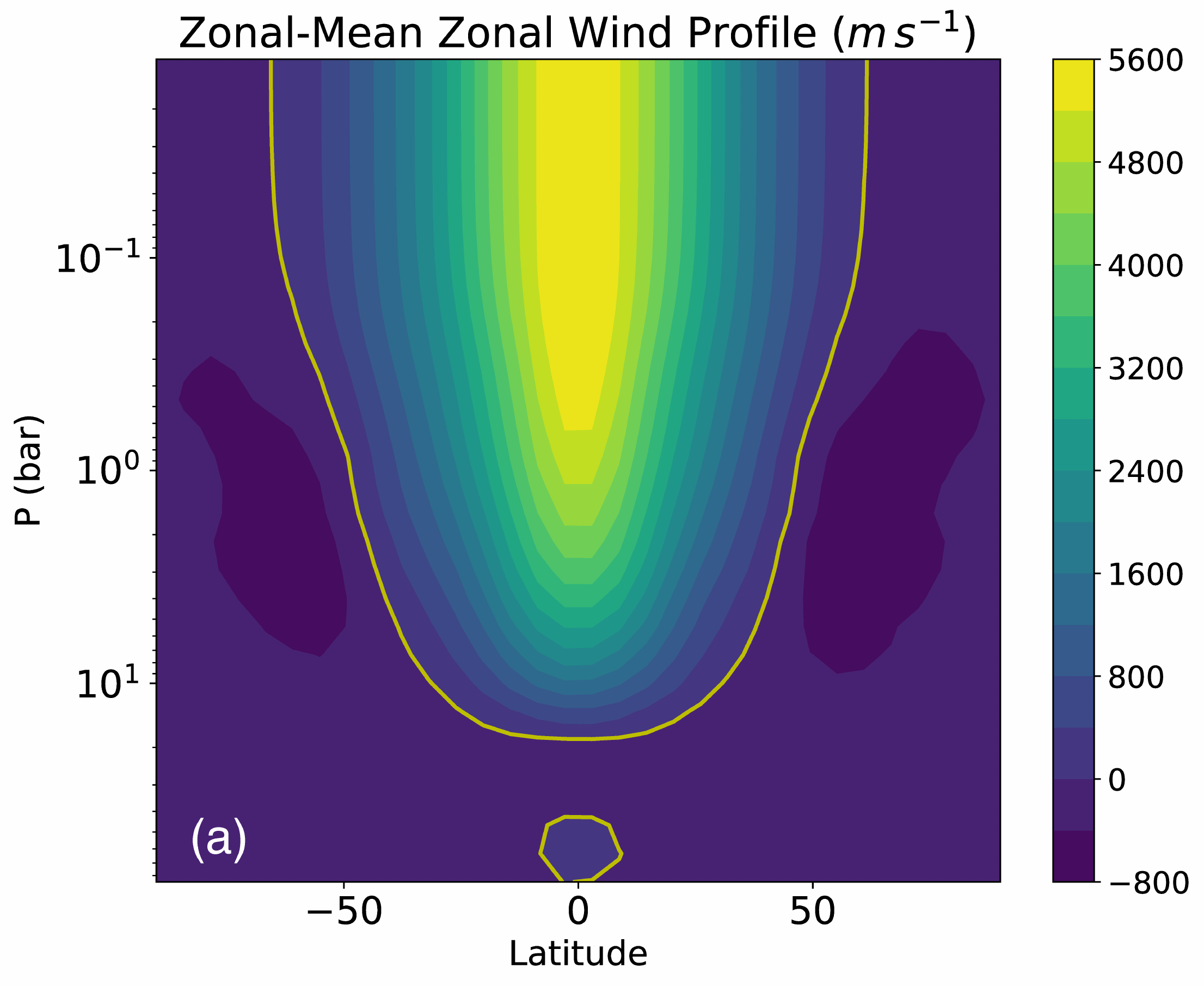}
	\includegraphics[width=\columnwidth]{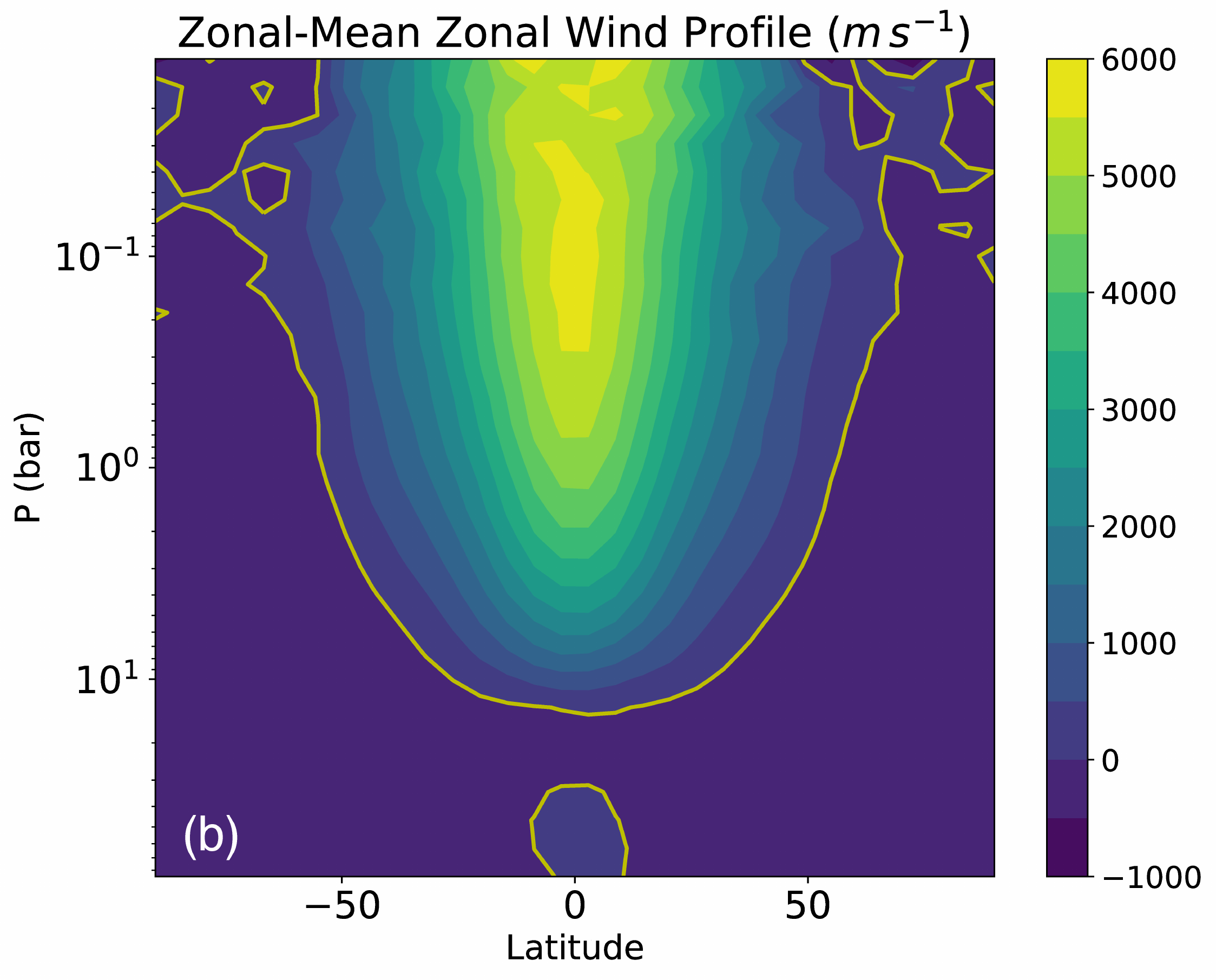}
    \caption{Zonally-averaged profile of the zonal wind in the HD209458b simulation with (a) and without (b) static imposed vertical transport. The zonal profile in (a) is mostly consistent with previously published results for this specific hot Jupiter \citep[e.g.][]{2012ApJ...750...96R}, but comparison with the profile in (b) for a similar model without any vertical transport shows some noticeable differences. Upper wind speeds have been somewhat reduced and vertically homogenized by momentum transport, especially higher up.}
    \label{fig:example_figure}
\end{figure}

Figure~1 shows the zonally-averaged zonal wind profile  in (a) our simulation with static imposed turbulent vertical transport according to Equation~(\ref{eq:transport}) and (b) a reference simulation without imposed vertical transport. The wind pattern that emerges is largely consistent with previously published wind profiles. This establishes that while turbulent momentum transport acts to reduce vertical shear in this simulation, it does not do so strongly enough to overcome the wind shear that is continuously established by differential radiative forcing.\footnote{Whether this result carries over to other exoplanets, especially hotter ones, is unclear and should be investigated explicitely.} Differences between panels (a) and (b) are noticeable, however. Wind speeds are somewhat reduced and vertically homogenized in the presence of strong vertical transport, especially in the 10-100 mbar region.

Taking the atmospheric patterns in these simulations at face value, we can run a variety of diagnostics on the atmospheric flow. In particular, we can directly compute values for the shear rate, the Brunt-Vaissala frequency, the Richardson number and the thermal diffusivity coefficient everywhere in the atmospheric flow.\footnote{We use second-order centered finite-differences to compute gradient quantities when needed}. The range of values we observe in the simulation with static imposed vertical transport are compiled in Tables~1 and~2 for conciseness.

\begin{table}
	\centering
	\caption{Range of values for $S$ and $N$ in our simulation of HD209458b with imposed vertical transport}
	\label{tab:example_table}
	\begin{tabular}{lcr} 
		\hline
		Pressure (bar) & $S$ (Hz) & $N$ (Hz)\\
		\hline
		0.01 & $7 \times 10^{-9}$ -$2 \times 10^{-4}$ & $2.2 \times 10^{-3}$ -$3.7 \times 10^{-3}$ \\
		0.1 & $2 \times 10^{-6}$ -$3 \times 10^{-3}$ & $2 \times 10^{-3}$ -$2.9 \times 10^{-3}$ \\
		1 & $2 \times 10^{-6}$ -$7 \times 10^{-3}$ & $2 \times 10^{-3}$ -$2.5 \times 10^{-3}$ \\ 
		\hline
	\end{tabular}
\end{table}

\begin{table}
	\centering
	\caption{Range of values for $Ri$, $\xi$ and $\nu_{\rm eff}$ in our simulation of HD209458b with imposed vertical transport}
	\label{tab:example_table}
	\begin{tabular}{lccr} 
		\hline
		Pressure (bar) &  $\log_{10}(Ri)$ & $\log_{10}(\xi [{\rm cm^2 s^{-1}}])$ & $\log_{10}(\nu_{\rm eff} [{\rm cm^2 s^{-1}}])$\\
		\hline
		0.01 &  2.28-11.4 & 11.3-13.5 & 9-11.6\\
		0.1  & 0.95 - 6.1 &10.5 - 12.0 & 10.0 -11.5\\
		1  & 0.24 - 6.1 & 9.3 - 10.2 & 8.0 - 10.1\\
		\hline
	\end{tabular}
\end{table}

We can also evaluate the secular Richardson number, $Ri_s$, and the expected strength of turbulent vertical transport in the form of a  diagnostic $K_{zz}$ value. Figure~2 shows detailed maps of local $K_{zz}$ values computed on the 40 (panel a) and 860 (b) mbar surface levels in the simulation with static imposed vertical transport. $K_{zz}$  values vary by orders of magnitude on a given pressure level with a pattern that follows the temperature field, leading to distinct dayside and nightside vertical transport regimes high in the atmosphere.

\begin{figure}
	\includegraphics[width=\columnwidth]{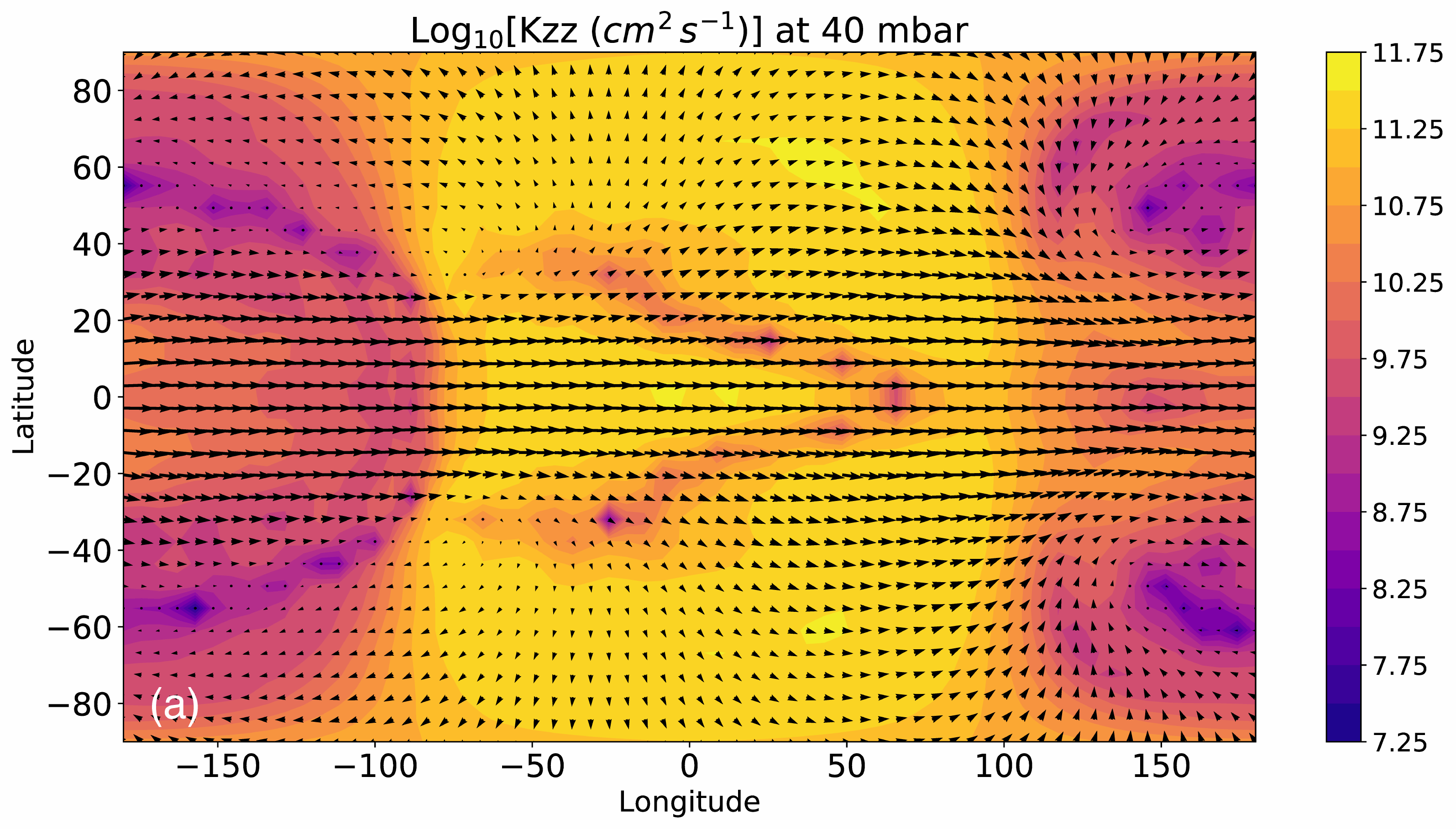}
	\includegraphics[width=\columnwidth]{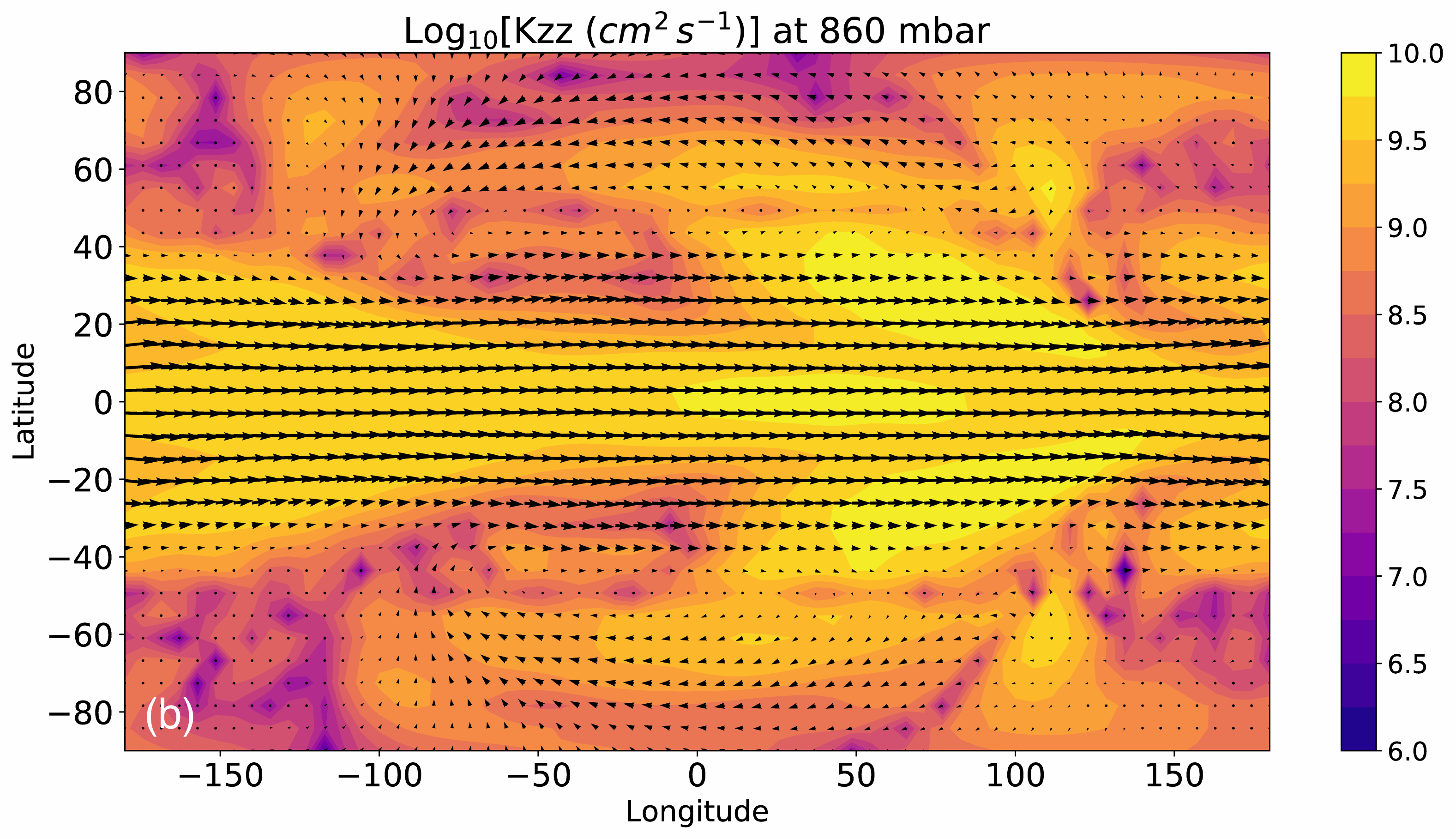}
    \caption{(a) Map of the diagnostic vertical mixing coefficient $K_{zz}$  calculated from the simulated flow on the 40 mbar pressure level in the simulation of HD209458b with static imposed vertical mixing. The map is centered on the sub-stellar point. The hotter dayside is the site of stronger mixing than the cooler night side, but only by about 2 orders of magnitudes.  (b) Same map as (a), but on the 860 mbar pressure level.}
    \label{fig:example_figure}
\end{figure}

\begin{figure}
	\includegraphics[width=\columnwidth]{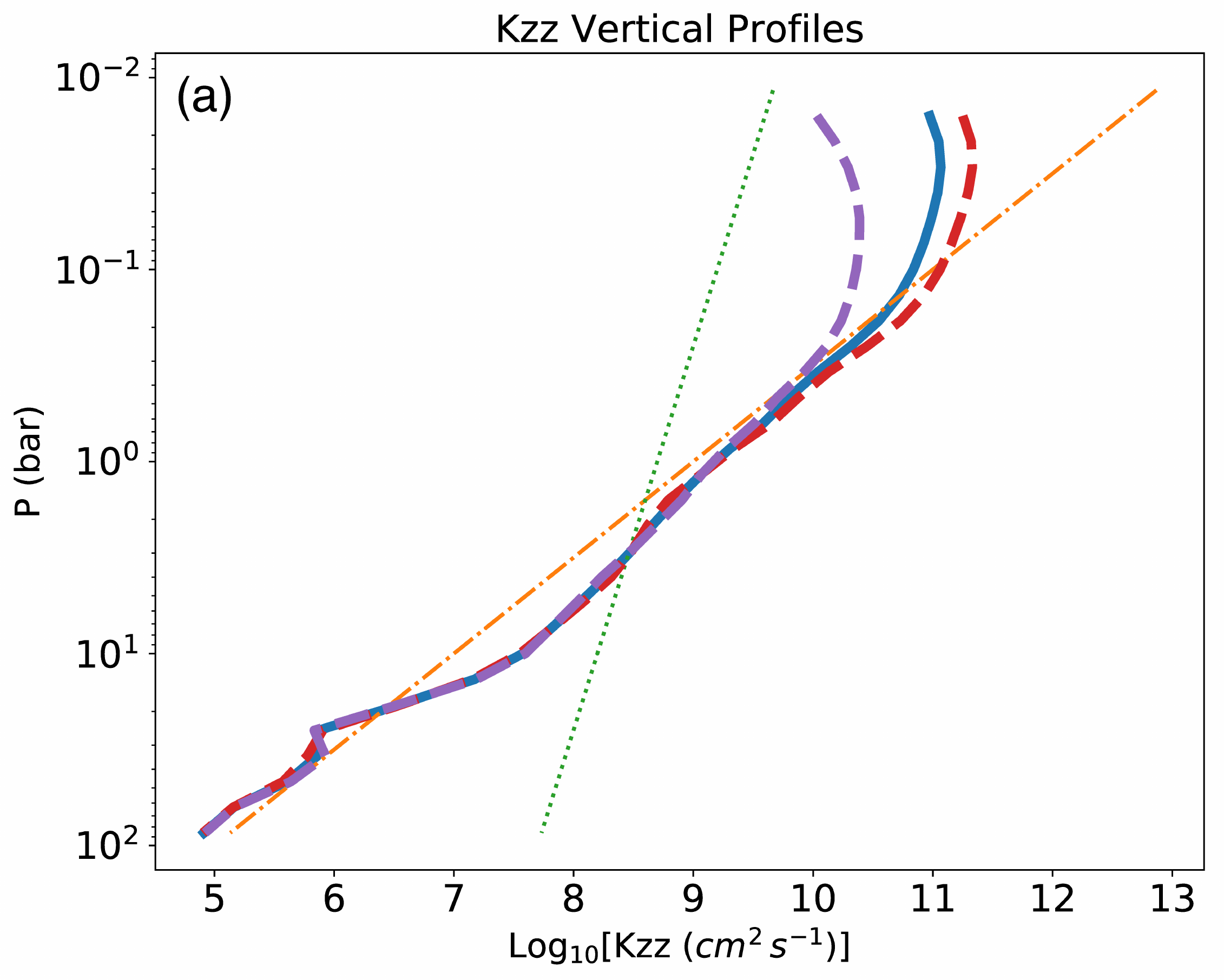}
	\includegraphics[width=\columnwidth]{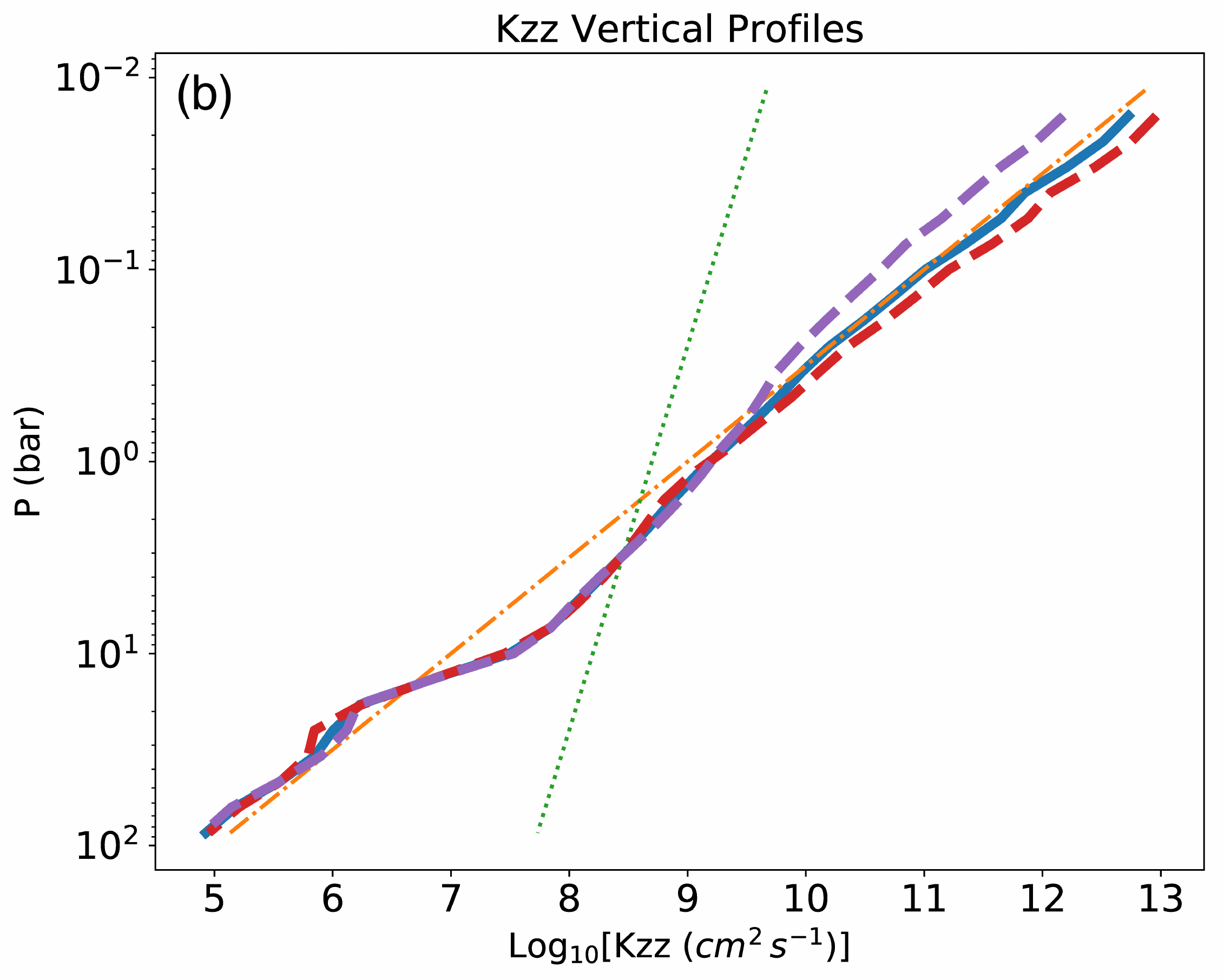}
    \caption{Vertical profiles of  $K_{zz}$, the diagnostic vertical mixing coefficient calculated from the simulated flow, in the cases with (panel a) and without (b) static-imposed vertical transport. In each panel, the solid blue line shows the profile inferred from the simulation of HD209458b, once $K_{zz}$ is averaged over pressure levels. The red and blue dashed show day-only and night-only averages. The dash-dotted orange line shows the static imposed vertical transport profile prescribed in the simulation ($\propto P^{-2}$) with imposed vertical transport. The green dotted line shows the mixing profile ($\propto P^{-{1/2}}$) advocated by Parmentier et al. (2013).
    }
    \label{fig:example_figure}
\end{figure}

Figure~3 shows vertical profiles of horizontally-averaged $K_{zz}$ values in the simulations with (panel a) and without (b) static-imposed vertical transport. In each panel, the solid blue line shows the diagnostic $K_{zz}$ profile directly computed from the simulation, after averaging over each pressure level. The red and blue dashed lines show day-only and night-only averages. The dash-dotted orange line shows the static vertical profile  ($\propto P^{-2}$)  imposed in one of the two simulations. The green dotted line shows the profile ($\propto P^{-{1/2}}$) advocated by \cite{2013A&A...558A..91P}. 

The good agreement between the profile of $K_{zz}$ shown as an orange dash-dotted line (following Eq.~13) and the profile calculated from the reference simulation without imposed transport (solid blue line in Fig.~3b) confirms that Eq.~13 is a good zeroth-order approximation for $K_{zz}$ values in the atmosphere of HD209458b. On the other hand, turn-overs in the diagnostic $K_{zz}$ profiles computed from the simulated flow with static imposed vertical transport (solid and dashed lines in Fig.~3a) show that strong vertical transport of momentum in the upper simulated atmosphere can act to reduce the wind shear present, sufficiently so that the resulting diagnostic $K_{zz}$ profiles are 1-2 orders of magnitude lower than in the simulation without imposed vertical transport.

This is a manifestation of the self-limiting nature of shear instabilities, where in the present case shear is continuously driven by differential radiative forcing and suppressed by the turbulent vertical transport that it generates. Since the diagnostic $K_{zz}$ profiles in Fig.~3a are found to be weaker than the static imposed vertical transport profile (orange dash-dotted line), the former can probably be used as lower-limits to the magnitude of transport that one would obtain in self-consistent simulations (with a transport magnitude continuously adjusted to what the atmospheric shear flow dictates). We also note that the diagnostic $K_{zz}$ profiles shown in Fig.~3a do not violate the limiting strength $K_{zz} \ll S H_p^2$, which is $ \sim 10^{12}$-$ 10^{13}$~cm$^{2}$s$^{-1}$ for this simulated flow.

Therefore, Figure~3a establishes that turbulent vertical transport driven by double-diffusive shear instabilities is strong and dominates over other sources of vertical mixing previously advocated for hot Jupiter atmospheres \citep{2013A&A...558A..91P} above approximately the 1~bar level. The strongest turbulent vertical transport occurs on the dayside but even the planet's nightside is the site of vigorous mixing above the 1~bar level, up to the thermal photosphere.

\section{Conclusions}

The magnitude of vertical transport and mixing in hot Jupiter atmospheres influences the disequilibrium chemistry of molecular species \citep[e.g.][]{2018ApJ...866....2Z} and determines the nature and extent of cloud decks maintained against the gravitational settling of condensates \citep[e.g.][]{2018ApJ...860...18P, 2018arXiv181203793H}.

\cite{2009ApJ...699.1487S} have estimated that values of $K_{zz} \geqslant 10^{11}~{\rm cm^2 ~s^{-1}}$ are needed to maintain particles of size $10\mu$m aloft on the dayside of hot Jupiters. Similarly, \cite{2013A&A...558A..91P} have estimated that values of $K_{zz} \geqslant 1.5 \times 10^{10}~{\rm cm^2 ~s^{-1}}$ are required to maintain particles of size $10\mu$m aloft across the nightside of hot Jupiters. These criteria are both met by the day- and night-averaged profiles shown in Fig.~3a (red and violet dashed lines, respectively). {This suggests that diffusive shear instabilities offer a promising solution to the need for strong vertical mixing emphasized by \cite{2009ApJ...699.1487S}  and \cite{2013A&A...558A..91P}.
Detailed models will be needed to address this issue quantitatively and establish whether the resulting mixing carries over to mbar pressure levels, as may be needed in some cases.}

Stronger mixing favors the maintenance of massive cloud decks and larger particle sizes  \citep[e.g.][]{2018ApJ...860...18P, 2018arXiv181203793H}. Our results indicate that, while the nightsides of hot Jupiters exhibit reduced vertical mixing relative to their daysides, nightside mixing is still vigorous enough to facilitate the maintenance of substantial nocturnal clouds, at least on  planets like HD209458b. Assuming this property carries over to other hot Jupiters, it could provide a key ingredient to explain recent results pointing to prevalent nightside clouds in the hot Jupiter population \citep{2018arXiv180809575B, 2018arXiv180900002K}.

While encouraging, it is also clear that our results on the role of double-diffusive shear instabilities remain of an exploratory nature. At a more fundamental fluid level, we would like to understand how double-diffusive shear instabilities develop, saturate and effectively transport both passive (condensates) and  active (momentum) flow attributes in a hot exoplanet atmosphere. In particular, momentum may be transported differently than passive tracers. These instabilities also abruptly cease to operate at/above the thermal photosphere, an aspect that deserves particular attention if one is interested in the observational signatures of hot Jupiters.

One would also like to gain a broader understanding of the range of applicability of these instabilities, from warm mini-Neptunes to the hottest hot Jupiters. This may require a fairly extensive and numerically expensive survey of global circulation models of the type presented here for HD209458b. On the hottest planets, turbulent vertical transport could compete with magnetic drag to reduce wind speeds at/below thermal photospheric levels. On the coolest planets where the process still operates, it might be restricted to their daysides, with much reduced mixing on the nightsides.

\section*{Acknowledgements}
The author is grateful to Jeremy Goodman for comments on an early version of this manuscript. KM is supported by the National Science and Engineering Research
Council of Canada. This work has made extensive use of the following software packages: {\tt matplotlib}.

We would furthermore like to acknowledge that our work was performed on land traditionally inhabited by the Wendat, the Anishnaabeg, Haudenosaunee, Metis, and the Mississaugas of the New Credit First Nation.




\bibliographystyle{mnras}
\bibliography{turb_mix} 








\bsp	
\label{lastpage}
\end{document}